\documentclass[10pt, conference, letterpaper]{IEEEtran}
\usepackage{caption}
\usepackage{etoolbox}
\newcounter{breakablesubsubsection}[section]
\renewcommand{\thebreakablesubsubsection}{\arabic{breakablesubsubsection})}

\newcommand{\breakableSubsubsection}[1]{%
  \par\vspace{0.8em}%
  \refstepcounter{breakablesubsubsection}%
  \noindent\textbf{\thebreakablesubsubsection\quad #1}\par\nopagebreak[1]\vspace{0.5em}%
}
\makeatletter
\patchcmd{\@startsection}
  {\@afterindenttrue \@afterheading}
  {\@afterindenttrue}
  {}{}
\makeatother

\usepackage{comment}
\usepackage[utf8]{inputenc}
\usepackage{tensor}
\usepackage[cmex10]{amsmath} 
\usepackage{physics}
\usepackage{amssymb,amsfonts}
\usepackage{mathtools,physics, bbm}

\usepackage{amsthm}
\newtheorem{theorem}{Theorem}

\newtheorem*{Game*}{Game}
\newtheorem*{TokGen*}{TokenGeneration Phase}
\newtheorem*{TokVer*}{TokenVerification Phase}
\newtheorem*{Setup*}{Setup}
\newtheorem*{Query*}{Query}
\newtheorem*{Challenge*}{Challenge}
\newtheorem*{Guess*}{Guess}
\usepackage{breqn}
\theoremstyle{remark}
\newtheorem{remark}[theorem]{Remark}
\usepackage{tikz}
\usetikzlibrary{quantikz}
\DeclareCaptionType{mycapequ}[][List of equations]
\usepackage{graphicx}

\usepackage{tikz}
\usetikzlibrary{quantikz}

\usepackage{graphicx}
\usepackage{url}
\usepackage{cite}
\usepackage{hyperref}
\hypersetup{
    hidelinks = true
}
\usepackage[capitalize]{cleveref}
\begin{document}

\title{Secure Hybrid Key Growing via Coherence Witnessing and Bipartite Encoding}

\author{
\IEEEauthorblockN{Pol Julià Farré$^*$, Chris Aaron Schneider  \& Christian Deppe}
\IEEEauthorblockA{Technische Universität Braunschweig, Braunschweig, Germany\\}

$^*$Corresponding author: pol.julia-farre@tu-bs.de
}

\maketitle

\begin{abstract}
We propose a novel Hybrid Key Growing (HKG) protocol based on quantum principles and a classical physical-layer assumption. We simultaneously exploit the quantum photon-number and photon-time-bin Degrees of Freedom (DoFs), effectively doubling the bit-per-pulse rate compared to conventional Quantum Key Growing (QKG) schemes. Our protocol integrates entity authentication, and is designed for practical implementation by avoiding reliance on single-photon sources or detectors. By incorporating prior knowledge about the quantum channel, the scheme actively mitigates noise effects, making it suitable for real-world conditions. Under certain assumptions on experimental efficiencies, our approach also promises an increased key generation rate in bits per second. Our simulation results display, first, expected outcomes to gain assurance about the correctness of our implementation and, second, relevant dependencies that showcase desirable properties of our scheme in regimes of low photon loss and dephasing. In particular, within such regimes, our encoding scheme reduces the Quantum Bit Error Rate (QBER) while preserving the ability to detect eavesdropping and identity-forgery attempts.
\end{abstract}


\section{Introduction}
\label{introduction}
Quantum Key Growing (QKG)~\cite{growing_1, growing_5, growing_6} is a cryptographic process that allows two parties—commonly referred to as Alice and Bob—to expand a small, initially shared secret key by exploiting the principles of quantum mechanics. More broadly, Quantum Key Distribution (QKD) refers to schemes that enable two parties to establish a shared secret key, even in the absence of an initial preshared secret. However, since QKD protocols often require a preshared key to ensure entity authentication, the term \textit{Quantum Key Growing} more accurately captures the underlying process in most practical implementations.
Current QKG protocols face several stringent limitations, some of which may be inherent to the underlying physical systems~\cite{limitations_QKD_1, limitations_QKD_2}. A notable example arises in prepare-and-measure-based QKG~\cite{prepare_and_measure_1, prepare_and_measure_2}, which typically relies on quantum communication through optical fibers~\cite{optical_fibers_1, optical_fibers_2}. These channels suffer from exponential signal attenuation with increasing distance, significantly limiting the achievable transmission range. To address this issue, satellite-based quantum links have been proposed as a scalable alternative~\cite{satellite_1, satellite_2}. In contrast to fiber links, satellite channels exhibit signal loss that scales only quadratically with distance, making them more suitable for long-range quantum communication~\cite{satellite_overview, satellite_cv}. However, their performance remains highly susceptible to atmospheric effects~\cite{atmospheric}, including diurnal variations~\cite{night_satellite, key_less_alternative}, which in turn complicate reliable simulation and prediction~\cite{hard_to_model_satellite}.
Quantum relays offer a practical solution to the range limitation in QKG by forwarding signals through intermediate nodes~\cite{relays_1, relays_2}. However, this approach requires complete trust in each relay node, which is generally undesirable.
 In contrast, quantum repeaters~\cite{q_repeaters_seminal, inside_q_repeaters,Loock2020} enable long-distance, entanglement-based QKG~\cite{entanglement_based_1, entanglement_based_2} without the need to trust intermediate nodes. Despite the fact that their practical deployment remains technologically challenging~\cite{repeaters_perspectives}, the development of protocols designed for repeater-based architectures ensures a smoother transition to a future in which quantum and classical communication coexist~\cite{6g1, 6g2}.
In parallel, prepare-and-measure QKG protocols remain highly relevant due to their technological maturity and wide range of practical applications. These include secure short-distance communication (up to a few hundred kilometers) and scenarios in which the use of trusted quantum relays is considered acceptable.

\subsection*{Our contributions}
We propose a novel Hybrid Key Growing (HKG) method that combines quantum principles—drawing from prepare-and-measure schemes—with a single classical assumption: that eavesdropping on the quantum channel causes a detectable delay. Our protocol leverages the main strength of one of the most practical prepare-and-measure QKG variants: the decoy-state technique~\cite{prepare_and_measure_2, decoy_bs_attack, decoy_realization}. Like the decoy-state approach, our protocol does not rely on ideal single-photon sources or detectors, thereby improving resilience to moderate photon loss and enhancing practical security. A central advantage of our method is that it eliminates the basis reconciliation (key-sifting) step found in all conventional QKG protocols, effectively doubling the bit-per-pulse key generation rate.
Existing schemes that also utilize multiple photonic Degrees of Freedom (DoFs)~\cite{Other_q_multiplexing} can achieve similar benefits, but typically incur a higher classical communication overhead. Our protocol, by contrast, adds complexity primarily at the transmitter and receiver hardware level, leaving its impact on the overall bit-per-second rate uncertain. Nevertheless, if this complexity does not significantly hinder throughput, the resulting gains could be highly valuable in latency-sensitive applications such as Digital Twin (DT) systems~\cite{DT, survey_DT}, where the rapid establishment and renewal of shared secret keys is essential in security-sensitive applications.
Our HKG protocol introduces a novel approach to handling quantum noise by leveraging full knowledge of the underlying two-qubit quantum channel—a reasonable assumption in many practical scenarios~\cite{p_tomography_1, p_tomography_2}. We optimize $L$, a defining parameter of our protocol,  to maximize performance under given noise conditions. Our simulations reveal that, under low-noise conditions, our encoding strategy decreases the Quantum Bit Error Rate (QBER) without compromising security.

\section{Preliminaries}
This section outlines the key mathematical tools and physical concepts used in this work. We present the optical-fiber channel model and introduce the photon-number encoding and coherence-witnessing techniques that underpin our protocol.

\subsection{Optical fibers: loss, dephasing and dark counts}

One of the cornerstones of quantum communication are fiber-based links, which, among other advantages, enjoy low costs by leveraging an extensive existing infrastructure \cite{recycle_fibers}, and low latency. However, they doubly suffer from exponential, with respect to the transmission distance $d$, signal degradation. That is, on the one hand, the optical-fiber-induced photon loss channel $\Xi_\eta$ is described via the Kraus operators
\begin{equation}
    K_{\eta, k} = \sum_{\substack{M \geq k}} \sqrt{p_{\eta}(k, M)} \, |M-k\rangle \langle M|, \hspace{0.2cm} \forall k \in  \mathbb{N},
    \label{loss}
\end{equation}
where the underlying Hilbert space is referred to the so-called Fock basis and  $p_{\eta}(k, M) := \binom{M}{k} (1 - \eta)^k \eta^{M-k}$,  with $\eta = 10^{-\frac{d \alpha }{10}}$ and $\alpha \approx 2\cdot 10^{-1}\hspace{0.1cm} \frac{\mathrm{dB}}{\mathrm{km}}$ \cite{repeaters_perspectives}. On the other hand, the dephasing channel $\Lambda_{\gamma}$  can be described as \cite{loss_dephasing}
\begin{equation}
  \bra{r}  \Lambda_\gamma \big( \rho \big) \ket{t} = f_\gamma(r-t)\bra{r}  \rho \ket{t}, \hspace{0.1cm}
\forall r, \hspace{0.05cm}t \in \mathbb{N}, \label{dephase}
\end{equation}
where $\rho$ is a general state, $f_\gamma(x) := e^{-\frac{\gamma}{2}x^2}$, and the dephasing parameter $\gamma$ scales linearly with time \cite{dephasing_time} and, hence,  with $d$.   
\begin{remark}
     Photon-loss and dephasing channels  commute \cite{loss_dephasing}.
\end{remark}  
Dark counts, are the last phenomenon considered within our model. We take them into account by lastly applying, to the sent states, a channel $\Gamma_\lambda$ described by the Kraus operators
\begin{equation}
    L_{\lambda,k} = \sum_{M=0}^{\infty} \sqrt{p_{\lambda}(k)} \ket{M+k}\bra{M}, \hspace{0.2cm} \forall k \in \mathbb{N},
\end{equation}
where $p_{\lambda}(k) := \frac{\lambda^k e^{-\lambda}}{k!}$, with $\lambda \approx 10^{-3}$ \cite{dark_count_rate} being the average number of dark counts within the selected time window.

\subsection{Bosonic encoding: a way out to moderate photon loss}

\label{bosonic_codes}

We introduce the rotation-symmetric bosonic codes \cite{bosonic_rotation} by defining the basis of a  logical  qubit in the following manner
\begin{equation}
     \ket{0_{N, \Theta}} = \frac{1}{\sqrt{\mathcal{N}_1}}\sum_{m = 0}^{2N-1} e^{i\frac{m\pi}{N}\hat{n}}\ket{\Theta} \hspace{0.2cm} \mathrm{and}
\end{equation}
\begin{equation}
    \ket{1_{N, \Theta}} = \frac{1}{\sqrt{\mathcal{N}_2}}\sum_{m = 0}^{2N-1} (-1)^m e^{i\frac{m\pi}{N}\hat{n}}\ket{\Theta},
\end{equation}
where $\mathcal{N}_1$ and $\mathcal{N}_2$ are normalization constants, $\hat{n}$ is the number operator,  and $\ket{\Theta}$ must fulfill
\begin{equation}
    \exists r, t \in \mathbb{N} \hspace{0.2cm} \text{s.t.} \hspace{0.2cm}
    \left\langle 2rN \middle| \Theta \right\rangle \neq 0 
   \hspace{0.2cm} \text{and} \hspace{0.2cm}
    \left\langle (2t+1)N \middle| \Theta \right\rangle \neq 0.
\end{equation}
By fixing $\ket{\Theta} = \frac{1}{\sqrt{2}} \big( \ket{N} + \ket{2N} \big) := \ket{\tilde{\Theta}}$, we obtain 

\begin{equation}
     \ket{ 0_{N, \tilde{\Theta}}} =   \ket{2N} := \ket{0}_N   \hspace{0.2cm} \mathrm{and}
\end{equation}
\begin{equation}
    \ket{1_{N, \tilde{\Theta}}} =  \ket{N} := \ket{1}_N ,
\end{equation}
leading to an enhanced readout distinguishability by assigning a ``$0$" to all photon counts $m$ fulfilling 
\begin{equation}
   \small
    \frac{\sum_{\substack{i \leq 2N, j \\ \text{s.t.} \ 2N - i + j = m}} p_{\eta}(i, 2N) p_{\lambda}(j)}{\sum_{\substack{i\leq 2N, j \\ \text{s.t.} \ 2N - i + j = m}} p_{\eta}(i, 2N) p_{\lambda}(j) + \sum_{\substack{i \leq N, j \\ \text{s.t.} \ N - i + j = m}} p_{\eta}(i, N) p_{\lambda}(j)} \geq
    \label{rule2}
\end{equation}
\begin{equation*}
    \small 
    \frac{\sum_{\substack{i\leq N, j \\ \text{s.t.} \ N - i + j = m}} p_{\eta}(i, N) p_{\lambda}(j)}{\sum_{\substack{i\leq N, j \\ \text{s.t.} \ N - i + j = m}} p_{\eta}(i, N) p_{\lambda}(j) + \sum_{\substack{i\leq 2N, j \\ \text{s.t.} \ 2N - i + j = m}} p_{\eta}(i, 2N) p_{\lambda}(j)},
\end{equation*}
and a ``$1$" otherwise.

\subsection{Coherence witness}

Given  the dephasing operation $\Delta$\cite{coherence_witness} and any Hermitian operator $\sigma$, we define a coherence witness  as the observable

\begin{equation}
    \hat{W}_\sigma := \Delta(\sigma) - \sigma.
\end{equation}
As shown in \cite{coherence_witness}, if  the expectation of $\hat{W}_\sigma$ over a certain state is negative, then that state must exhibit quantum coherence, or as  often also termed: quantum superposition.

\begin{remark}
    If $\sigma$ is a density matrix, the expectation of $\hat{W}_\sigma$ over an ensemble of quantum states can be computed using the swap test \cite{swap_test, swap_test_experiment, swap_proof_2}.
\end{remark}

\section{A new HKG protocol}

In this section, we describe, in five phases, a Hybrid Key Growing (HKG)  protocol between two parties: Alice and Bob. The first run of the protocol is exactly prescribed within the five shown phases, while further rounds require slight modifications that are detailed in the 5th phase. 

\breakableSubsubsection{Quantum-bipartite key encoding}

In the first phase, Alice, the sender, generates a bitstring $\mathbf{K}$ uniformly at random of length $\beta = \nu \omega$, with $\omega,$ $\nu \in \mathbb{N}$ and with $\nu > \omega$. Subsequently, she encodes each bit of $\mathbf{K}$ as indicated by another bitstring $\mathbf{F}$, also generated uniformly at random, of the same length as $\mathbf{K}$, and preshared with Bob: If a certain bit of $\mathbf{F}$ is a ``$0$" Alice encodes the corresponding bit $b_0$ of $\mathbf{K}$ as  
\begin{equation}
    \ket{ b_0}_N \otimes \frac{1}{\sqrt{2}} (\ket{0}_{\mathrm{\mathrm{time}\text{-}\mathrm{bin}}} + (-1)^{\xi}\ket{1}_{\mathrm{\mathrm{time}\text{-}\mathrm{bin}}}),
    \label{bosonic_qubits}
\end{equation}
where kets with the subscript \textit{time-bin} indicate qubit states belonging to the Hilbert space spanned by two distinct and orthogonal pulse time bins, and $\xi$ is a uniformly random bit Alice needs to generate uniformly at random at each preparation.  Instead, if a certain bit of $\mathbf{F}$ is a ``$1$", Alice encodes the corresponding bit $b_1$ of $\mathbf{K}$ as

\begin{equation}
     \frac{1}{\sqrt{2}} (\ket{0}_N  +(-1)^{\xi} \mid 1\rangle_N) \otimes \ket{b_1}_{\mathrm{\mathrm{time}\text{-}\mathrm{bin}}}.
    \label{time-bin_qubits}
\end{equation}

\breakableSubsubsection{Decoding and coherence-witness estimation}

Bob receives the sent  pulses, and performs different operations on them, depending on their type.  He measures those prepared  as in Eq. \eqref{bosonic_qubits}  in the Fock basis \cite{fock_definition} to extract, for each of them, a secret bit $\tilde{b}_0$. Given low quantum noise, the bit $\tilde{b}_0$  likely coincides with the bit $b_0$ encoded by Alice. Moreover, the time-bin Degree of Freedom (DoF) of such incoming pulses is coupled with another reference system prepared as
\begin{equation}
        \ket{\phi} := \frac{1}{\sqrt{2}}(\ket{0}_{\mathrm{\mathrm{time}\text{-}\mathrm{bin}}} + \ket{1}_{\mathrm{time}\text{-}\mathrm{bin}})
        \label{time_bin_reference}
\end{equation}
and an ancillary qubit used to perform the swap test between the received pulse and $\ket{\phi}$. These swap tests are later used to average the coherence witness
\begin{equation}
       \hat{W}_{\mathrm{time\text{-}bin}} :=  \hat{W}_{\ket{\phi}\bra{\phi}}
\end{equation}
over certain input states. On the other hand, Bob measures the time-bin of the incoming pulses prepared as in Eq. \eqref{time-bin_qubits} to obtain, for each pulse, an  estimated bit $\tilde{b}_1$. Being $L$  a natural parameter  chosen after channel characterization (see Section \ref{methods}), Bob interacts with the photon-number property of each incoming pulse by coupling it with another reference system 
\begin{equation}
        \ket{\psi}(L) := \frac{1}{\sqrt{2}}(\ket{L + N} + \ket{L})
        \label{photon_number_reference}
\end{equation}
and an ancillary qubit used to perform the swap test between the state of the received pulse and $\ket{\psi}(L)$. These swap tests are later used to average the coherence witness
\begin{equation}
       \hat{W}_{\mathrm{Fock}}(L) :=  \hat{W}_{\ket{\psi} \bra{\psi}(L)}
\end{equation}
over certain input states.

\breakableSubsubsection{Authentication decisions and eavesdropping check}
 Alice and Bob publicly compare a small fraction $\kappa$, discarded for later use, of the obtained raw key. A Quantum Bit Error Rate (QBER) smaller than a threshold $\mu < \frac{1}{4}$, defined during channel characterization, leads Alice to regard the receiver as legitimate. Otherwise, the protocol is aborted.

Under no protocol abortion, Alice publicly shares the choice of $\xi$ made in each pulse preparation. Bob estimates, only selecting those pulses sent with $\xi = 0$, the  mean value  of the two  coherence witnesses presented, defined as
\begin{equation}
    \langle \hat{W}_{\mathrm{time\text{-}bin}}\rangle_{\mathrm{selection_1}} \hspace{0.2cm} \mathrm{and} \hspace{0.2cm} \langle \hat{W}_{\mathrm{Fock}}(L)\rangle_{\mathrm{selection_2}},
\end{equation}
and computes the average $W(L)$ between them. An obtained value below a threshold $-\frac{1}{2}< \tau < 0$, defined during channel characterization, ensures both no eavesdropping and sender's legitimacy. Otherwise, the protocol is aborted.

\breakableSubsubsection{Key distillation}

Under no protocol abortion, the two parties proceed with the information reconciliation \cite{info_reconciliation} and privacy amplification \cite{priv_amplification} schemes, yielding a sifted, yet correct and secret, shared key. 

\begin{remark}
    Our HKG protocol does not require the commonly used bases-reconciliation step, thus reducing the classical communication overhead and approximately doubling, due to no bases mismatching, the bit-per-pulse key rate with respect to conventional QKG schemes.
\end{remark}

\breakableSubsubsection{$\mathbf{F}$ update and further-round indications}
\label{f_update}

$\mathbf{F}$ is divided into $\omega$ segments, each of $\nu$ bits by construction. Alice and Bob select, and discard for later use, $\omega$ of the secret shared bits generated. For bits  of value ``$1$", their corresponding segments in $\mathbf{F}$ are flipped.  Otherwise, i.e., for bits of value  ``$0$", the corresponding bit segments remain unchanged. Furthermore, in any $k$-th round, with $k>1$, Bob's authentication (see 3rd phase) takes place by him and Alice comparing $\omega$ sets of bits, each set belonging to one of the defined segments, and each set being of size $\kappa \nu$. Bob is  accepted if the QBER is smaller than the threshold $\mu$ in all $\omega$ instances and, otherwise, the protocol is aborted. Under no protocol abortion, Bob confirms Alice's identity by computing  $W$ for each corresponding set. Alice is accepted if $W<\tau$ in all $\omega$ instances and, otherwise, the protocol is aborted.
\section{Security analysis}
\label{sec_analysis}
 Correctness \cite{security_definition_1} of the final shared key trivially follows from positive authentication decisions together with a negative eavesdropping check. We focus  on  entity-authentication and key-secrecy constraints, relying, for the former, on the completeness-and-soundness framework \cite{authenticated_bb84}.

 \subsection{Authentication completeness}
 
  With regard to completeness, Bob is trivially accepted given no malicious parties and no quantum noise. In such a case, and setting $L = N$, Alice is also accepted, because Bob performs the swap test against a state that flags each coherence witness in all instances. Namely, ideally \cite{born_rule, wilde},

 \begin{equation}
     \langle \hat{W}_{\mathrm{time\text{-}bin}}\rangle_{\mathrm{selection_1}} =  -\frac{1}{2} \hspace{0.2cm} \mathrm{and} 
 \end{equation}
 \begin{equation}
    \small
    \langle \hat{W}_{\mathrm{Fock}}(N)\rangle_{\mathrm{selection_2}} = -\frac{1}{2}.
 \end{equation}
Moreover, given the continuity of both the expected QBER and $W$ with respect to quantum-state fidelity, an adequate choice of $\mu$ and $\tau$ ensures completeness  in  setups with low-noise conditions.

 \subsection{Authentication soundness}
 With regard to soundness, we divide our analysis into two parts, depending on the party that is aimed to be forged.

\subsubsection{Resistance against Bob-impersonation attacks}

Firstly, let us consider an adversary not knowing $\mathbf{F}$, and  performing  a straightforward man-in-the-middle attack \cite{straightforward}. For each incoming pulse, the probability of measuring the right DoF is  $\frac{1}{2}$, thus leading to an expected QBER lower bounded by $\frac{1}{4}> \mu$. Secondly,  notice that only eavesdropping the used quantum channel provides no information about $\mathbf{F}$. However, such an eavesdropping combined with the disclosure of the choices of each $\xi$ and the information reconciliation scheme—both of which are public by design—may compromise the secrecy of $\mathbf{F}$. Desirably, the 5th phase of our scheme ensures that, even with full knowledge of $\mathbf{F}$, the probability of correctly forging Bob is roughly equal to $\frac{1}{2^\omega}$. That is, an attacker knowing $\mathbf{F}$, prior to its update, and aiming to impersonate Bob, measures with a high probability $p = 1-\frac{1}{2^\omega}$ the wrong DoF for, at least, one complete subset of  $\nu$ pulses. Hence, the expected QBER for such a subset is $\frac{1}{2}> \mu$.

\subsubsection{Resistance against Alice-impersonation attacks}

Two conditions are needed to impersonate Alice: after generating $\beta$ bits, encoding each of them in a light pulse and sending them to Bob, both  the QBER and $W$ obtained must be low enough. This renders the attacker's encoding choices nontrivial. If only one condition was to be fulfilled, an adversary could double-encode a bitstring (to avoid a high QBER) or double-encode superposition states (to avoid a high value of $W$). We again first consider an adversary with no knowledge of $\mathbf{F}$. Each of the adversary's encoding guesses is successful with a probability $\frac{1}{2}$. Therefore, on average, half of the times in which Bob, under no forgery attempt, would perform the swap test against a state flagging a coherence witness, he does so with non-superposition states, thus substantially increasing $W$. Moreover, such an attack would induce an expected QBER of $\frac{1}{4}$. Finally, an attacker who knows $\mathbf{F}$, and makes a guess about the key update, encodes with high probability $p=1-\frac{1}{2^\omega}$ all $\nu$ bits of at least one of the $\omega$ underlying subsets in the wrong DoF. In this case, the QBER in at least one compared bit set would be $\frac{1}{2}$, and the value of $W$ in this set would drop to zero.

\subsection{Key secrecy}

As  stated above, $\mathbf{F}$ may be leaked to an eavesdropper harnessing public discussions and measurements on the sent stream of light pulses. This can compromise  key secrecy against the well-known intercept-resend \cite{intercept-resend} attack. However, we make an assumption about the physical layer: an intercept-resend attack would introduce substantial time delays on the light pulses, thus perturbing their time-bin DoF and thus increasing both the QBER and the value of $W$. This assumption, which relies on the specifics of the channel engineering and the adversary's classical capabilities, is what renders our protocol hybrid.

\section{Simulation}

\subsection{Methodology}
\label{methods}

The simulation methodology focuses in the optimization problem underlying our protocol definition. We  tune the integer parameter $L$  in order to minimize the  expectation of the coherence witness $\hat{W}_{\mathrm{Fock}}$ over the state 
\begin{equation}
    \theta_{N, \eta, \gamma, \lambda} = \Gamma_\lambda \Bigg( \Lambda_\gamma \bigg(  \Xi_\eta \Big(  \frac{1}{2} \sum_{i, j\in \{0, 1\}} \mid i\rangle_N {}_N\langle j\mid   \Big) \bigg)   \Bigg),
\end{equation}
i.e., that one describing the selection of incoming pulses that Bob measures for computing such an expectation. This leads to the definition of our cost function as\footnote{Notice that, in practice, small dark-count effects allow to truncate the infinite sum appearing in Eq. \eqref{inf_sum} into just a few terms.} 
\begin{equation}
    \chi_{N, \eta, \gamma, \lambda}(l) := 
    \langle \hat{W}_{\mathrm{Fock}} (l) \rangle_{\theta_{N, \eta, \gamma, \lambda}} = 
    \label{inf_sum}
\end{equation}
\begin{equation*} 
=-\frac{1}{2} \sum_{m = 0}^\infty  f_\gamma(N) \vartheta_{\eta, \lambda, \gamma}(N,m, l),
\end{equation*}
where
\begin{equation}
     \vartheta_{\eta, \lambda, \gamma}(x,y,z) :=
\end{equation}
\begin{equation*} 
\small
\begin{cases}
p_\lambda(y)\sqrt{p_\eta(x+y-z, 2x)p_\eta(x+y-z, x)} & \text{if }   x + y-z \geq 0 , \\
0 & \text{otherwise}.
\end{cases}.
\end{equation*}
Notice that minimizing $\chi_{N, \eta, \gamma, \lambda}(l)$ over $l$  yields the value of $L$ used at the implementation of the protocol. We propose to perform such a minimization via brute-fore search, given the low dimension of the domain space.

\subsection{Results}
\label{results}
In this section, we display the optimal value $L$ obtained for different values of $N$ and varying noise conditions. We argue why the obtained outcome is an expected one, which later serves  to robustly present and discuss relevant variable dependencies obtained with our numerical implementation.

\begin{figure}[h!]
    \centering
    \includegraphics[width=0.45\textwidth]{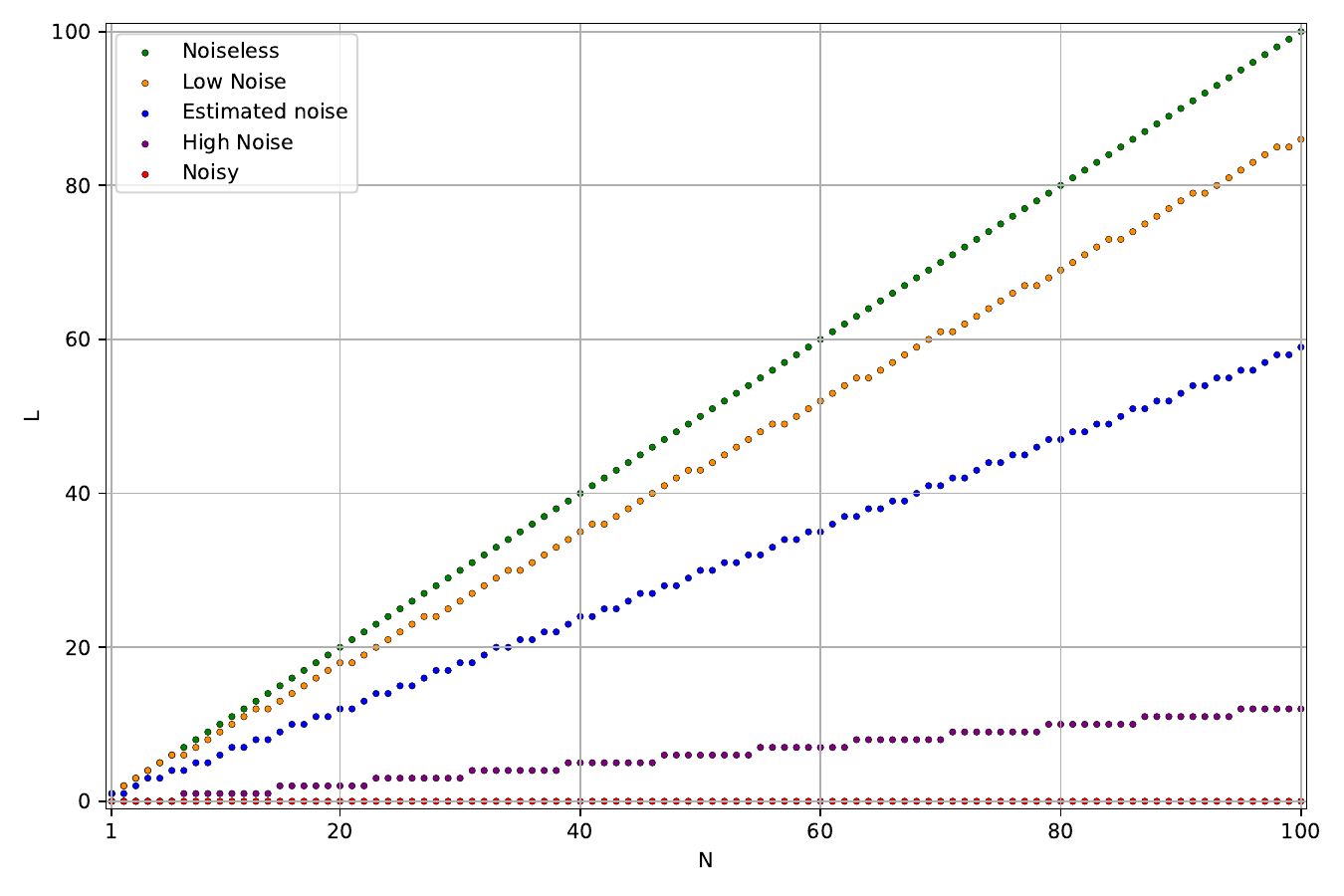}
    \caption{\small Optimal value $L$ (left vertical axis) and $\chi_{{N, \eta, \gamma, \lambda}} (L)$ (right vertical axis) in terms of $N$. Estimated-noise conditions are defined by setting $\eta = 7 \cdot 10^{-1}$ and $\lambda = 10^{-3}$, as reported, respectively,  in \cite{repeaters_perspectives} (for a distance of around $8$ km between parties), and \cite{dark_count_rate}. Because of lack of experimental evidence in the literature for  plausible estimations for the dephasing parameter, we set it to the arbitrary value $\gamma = 10^{-4}$, to obtain informative dependencies. Low-noise conditions are set by fixing $\eta = 9 \cdot 10^{-1}$, $\lambda = 10^{-3}$, and $\gamma = 10^{-4}$, while for high-noise conditions we shift to   $\eta = 3\cdot 10^{-1}$.}
    \label{fig_optimization}
\end{figure}

Fig. \ref{fig_optimization}, shows how, for a fixed $N$, the optimal value $L$ decreases with increasing noise regimes. In the noiseless case, $L$ coincides with $N$, i.e., with that value that maximizes the fidelity between the reference state $\ket{\psi}(L)$ and the state it is swapped-tested against assuming an ideal transmission. For higher noise conditions, instead, the value of $L$ gradually decreases until it vanishes for merely noisy data. This can also be understood by noticing that, when photon loss increases, new coherence terms appear in the Fock basis, of lower photon number and without disturbing the initial  photon-number difference between contributing terms, i.e., $N$.

\begin{figure}[h!]
    \centering
    \includegraphics[width=0.5\textwidth]{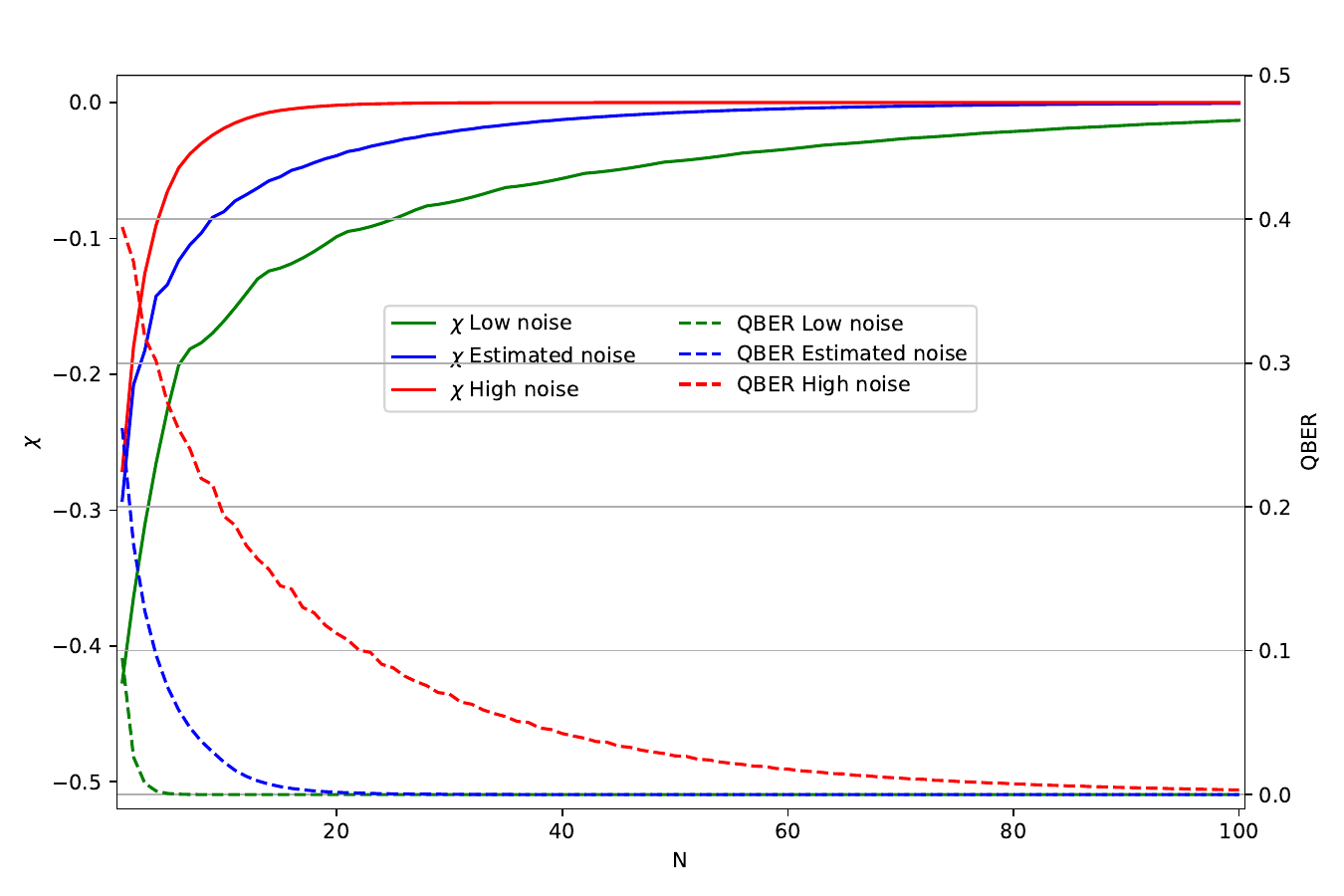}
    \caption{\small For various noise levels, the optimal $\chi$ (left axis) and expected QBER with our bosonic qubit encoding (right axis) are shown as a function of $N$. Noise characterizations match those in Fig. \ref{fig_optimization}.}
    \label{qber_and_cost}
\end{figure}

Fig. \ref{qber_and_cost} shows, for three different noise levels also studied in Fig. \ref{fig_optimization}, the dependency on $N$ for both the QBER obtained when reading out photon-number-encoded bits and the optimal value of $\chi$. Remarkably, for both low-noise conditions and our estimated noise impact for a distance between parties of around $8$ km, the QBER rapidly decreases with $N$ while $\chi$ increases at a  slower  pace. Thus, for the indicated noise conditions, and for certain moderate values of $N$, our proposal enlarges the generated raw key and lowers its QBER with respect to single-photon-based implementation, while remaining secure.

\section{Conclusions and further research}

Quantum Key Growing (QKG) is of critical importance in the current landscape of quantum communications, both due to its theoretical and experimental progress and its essential role in countering emerging quantum threats~\cite{shor, quantum_threat}. In this work, we have proposed a novel Hybrid Key Growing (HKG) protocol based on bipartite encoding, coherence witnessing, and a classical assumption on the physical layer. As shown in Sections~\ref{sec_analysis} and~\ref{results}, the protocol is compatible with real-world implementation under moderate noise conditions, assuming that our encoding-and-decoding scheme is plausible.

Although mature solutions such as decoy-state QKG remain highly effective and widely adopted, our approach explores alternative strategies with distinct operational principles. We believe that such a diversity is vital to advancing in the field. Future research will examine whether the adversarial model used in our analysis poses challenges to existing protocols and whether our method improves scalability in networked settings, e.g., in Hybrid Conference Key Growing.

Finally, as we became aware during the course of our research, we notice that generating and detecting states in the Fock basis remains a significant challenge with current technology. The recent work presented in \cite{delicate_Fock} offers a promising direction for overcoming these difficulties.

\section*{Code Availability}
\label{code}
All the  codes required for obtaining our simulation results  are publicly available at \cite{code_repo}.

\section*{Acknowledgements}

The authors thank Ángeles Vázquez-Castro, Pau Colomer, and Pere Munar-Vallespir for valuable discussions and proofreading. The authors  acknowledge financial support by the Federal Ministry of Research, Technology and Space of Germany in the program of “Souverän. Digital. Vernetzt”. Joint project 6G-life, project identification numbers: 16KISK002 and 16KISK263. CD and PJF were further supported under projects 16KISQ169, 16KIS2196, 16KISQ038, 16KISR038, 16KISQ0170, and 16KIS2234.


\end{document}